# Confining Metal-Halide Perovskites in Nanoporous Thin Films

Stepan Demchyshyn[1,2,3*], Janina Melanie Roemer,[4*] Heiko Groiß,[5,6] Herwig Heilbrunner,[1] Christoph Ulbricht,[1†] Dogukan Apaydin,[1] Anton Böhm,[4] Uta Rütt,[7] Florian Bertram,[7] Günter Hesser,[6] Markus Clark Scharber,[1] Niyazi Serdar Sariciftci,[1] Bert Nickel,[4,8] Siegfried Bauer,[2] Eric Daniel Głowacki,[9‡] Martin Kaltenbrunner[2,3‡]

[1] Linz Institute for Organic Solar Cells (LIOS), Physical Chemistry, Johannes Kepler University Linz, Altenbergerstraße 69, 4040 Linz, Austria

[2] Department of Soft Matter Physics, Johannes Kepler University Linz, Altenbergerstraße 69, 4040 Linz, Austria

[3] Linz Institute of Technology LIT, Altenbergerstraße 69, 4040 Linz, Austria

[4] Faculty of Physics & CeNS, Ludwig-Maximilians-University München, Geschwister-Scholl-Platz 1, 80539 München, Germany

[5] Christian Doppler Laboratory for Microscopic and Spectroscopic Material Characterisation, Johannes Kepler University Linz, Altenbergerstraße 69, 4040 Linz, Austria

[6] Center for Surface and Nanoanalytics, Johannes Kepler University Linz, Altenbergerstraße 69, 4040 Linz, Austria

[7] Deutsches Elektronen-Synchrotron (DESY) Photon Science, Notkestraße 85, 22603 Hamburg, Germany

[8] Nanosystems Initiative Munich, Schellingstraße 4, 80799 München, Germany

[9] Laboratory of Organic Electronics, Department of Science and Technology, Campus Norrköping, Linköpings Universitet, Bredgatan 33, SE-601 74 Norrköping, Sweden

[*] These authors contributed equally

[†] Present address: Institute for Polymeric Materials and Testing (IPMT), Johannes Kepler University Linz, Altenbergerstraße 69, 4040 Linz, Austria

[‡] Corresponding author. Email: eric.glowacki@liu.se (E.D.G.); martin.kaltenbrunner@jku.at (M.K.)

**Abstract**

Controlling size and shape of semiconducting nanocrystals advances nanoelectronics and photonics. Quantum confined, inexpensive, solution derived metal halide perovskites offer narrow band, color-pure emitters as integral parts of next-generation displays and optoelectronic devices. We use nanoporous silicon and alumina thin films as templates for the growth of perovskite nanocrystallites directly within device-relevant architectures without the use of colloidal stabilization. We find significantly blue shifted photoluminescence emission by reducing the pore size; normally infrared-emitting materials become visibly red, green-emitting materials cyan and blue. Confining perovskite nanocrystals within porous oxide thin films drastically increases photoluminescence stability as the templates auspiciously serve as encapsulation. We quantify the template-induced size of the perovskite crystals in nanoporous silicon with microfocus high-energy X-ray depth profiling in transmission geometry, verifying the growth of perovskite nanocrystals throughout the entire thickness of the nanoporous films. Low-voltage electroluminescent diodes with narrow, blue-shifted emission fabricated from nanocrystalline perovskites grown in embedded nanoporous alumina thin films substantiate our general concept for next generation photonic devices.

**Introduction**

Tuning the band gap of semiconductors via quantum size effects launched a technological revolution in optoelectronics, advancing solar cells (*1*, *2*), quantum dot light-emitting displays (*3*), and solid state lasers (*4*). Next generation devices seek to employ low-cost, easily processable semiconductors. A promising class of such materials are metal-halide perovskites (*5*), currently propelling research on emerging photovoltaics (*6–8*). Their narrow band emission permits very high color purity in light-emitting devices and vivid life-like displays paired with low-temperature processing through printing-compatible methods (*9–13*). The success of perovskites in light-emitting devices is conditional upon finding reliable strategies to tune the band gap while preserving good electrical transport. Thus far, color can be tuned chemically by mixed halide stoichiometry and by synthesis of colloidal nanoparticles (*14, 15*). Here we introduce a general strategy of confining perovskite nanocrystallites (less than 10 nm) directly within device-relevant solid state thin film formats without the use of colloidal stabilization. With nanoporous alumina (npAAO) or silicon (npSi) scaffolds serving as templates for the growth of perovskite nanocrystallites on transparent conductive oxides or silicon wafers, we achieve fine-tuning of the band gap across a wide color gamut from near infrared (NIR) to ultraviolet (UV). Confinement in npSi facilitates a ~50 nm hypsochromic shift from green to blue photoluminescence (PL) for cesium-bromide perovskite nanocrystals. By infiltrating npAAO templates on transparent conductive substrates, we fabricate perovskite light-emitting diodes (LEDs) that achieve blue-shifted narrow-band [full width at half maximum (FWHM), 17 nm] emission. Our demonstrations corroborate band gap engineering through confinement in nanoporous solids as a powerful tool to precisely control the emission wavelength of perovskite nanocrystals in next generation solution-derived photonic sources.

Quantum confinement was recognized as an auspicious feature in two-dimensional layered organic-inorganic hybrid perovskites in the pioneering work of Mitzi *et al*. (*5*, *16*, *17*) in the 1990s. Renewed interest in the alkylammonium lead halide (APbX$_3$) perovskites inspired elaboration of colloidal chemistry techniques, such as utilizing nanocrystalline capping (*18*, *19*), two step processes with PbI$_2$ nanocrystals as templates *(20)*, emulsion synthesis *(21)*, and reprecipitation (*22–24*) to achieve crystals featuring quantum size effects (*25*, *26*). Colloidal nanocrystals based on the all-inorganic CsPbX$_3$ system were developed by Protesescu *et al.* (*15*), which exhibit band gap tunability via halide composition *(14)* and quantum size effects *(15)*. The size dependency o the observed hypsochromic shifts can be well approximated by the confinement energy in a spherical potential well for the crystallite sizes above 6 nm *(15)*. In a ligand-free approach, nanoporous silica powder can be used as a template to achieve monodisperse lead halide perovskite crystals with quantum confinement (*27*, *28*). Here, the insulating nature of the silica precluded potential electronic device applications. Templating approaches based on mesoporous alumina to achieve nanoparticles *(29, 30)* were reported in geometries not allowing for electrical excitation.

**Results**

**Nanoscale reactors**

Here, we introduce solid-state nanoscale templating in a thin-film format to confine the growth of the perovskite directly within device-relevant architectures (Fig. 1A). This method yields perovskite nanocrystals with sizes of a few nanometers. We characterize nanocrystalline methylammonium lead trihalide (ncMAPbX$_3$), with chloride, bromide, and iodide (X = Cl, Br, I) perovskites, as well as nanocrystalline cesium lead tribromide (ncCsPbBr$_3$). We demonstrate the concept on two different nanoscaffold films, each with specific advantages. The first template comprises anodic aluminum oxide nanotubes (npAAO) with diameters of ~6-8 nm and lengths from 20 up to hundreds of nanometers. The second template is composed of electrochemically etched npSi layer tens of micrometers thick renowned for its 3D spongy network of nanopores with tunable size from 2 to 10 nm (*31*, *32*). Both nanoscale solid-state scaffolds are prepared using scalable, room temperature anodization (detailed in Materials and Methods). The infiltrated pores serve as nanoreactors, constraining the growth of the perovskite crystals that form from the solution-borne precursors. We imaged ~150-nm-thick npAAO cross sections using scanning transmission electron microscopy (STEM) (Fig. 1B). The projection of the STEM image resolves both a network of perovskite nanocrystals partially filling the nanopores and the surrounding AAO matrix. Here, the nucleation density (resulting in the apparent nucleation profile) is controllable via the concentration of the precursor solution (see the Supplementary Information). Exfoliated flakes of a npSi sponge filled with nanocrystals are shown in Fig. 1C. The perovskite nanocrystals appear dark (due to diffraction and mass contrast) in both matrices in these bright-field scanning transmission electron microscopy (BF-STEM) images. We find blue-shifted photoluminescence (PL) in all cases; nanocrystalline methylammonium lead triiodide (ncMAPbI$_3$) grown in npAAO blue shifts by 62 nm, while ncMAPbBr$_3$ and ncCsPbBr$_3$ shift by 14 and 16 nm, respectively (Fig. 1D). Blue shifts in the photoluminescence are common in organic conjugated molecules when isolated (*33*, *34*), for example in solution or in npAAO (*35–37*) and silica (*38*, *39*). In contrast, solid-state nanopore templating of nanocrystalline perovskites at sizes less than the exciton radius in the semiconductor *(40)* gives rise to strong quantum and dielectric confinement effects responsible for the observed blue shifted emission. The npSi, featuring smaller pores than npAAO, affords shifts of up to 151 nm for ncMAPbI$_3$ and of ca. 50 nm for ncMAPbBr$_3$ and ncCsPbBr$_3$, and typically 15 nm for ncMAPbCl$_3$ (Fig. 1E). These shifts correspond to significant band gap changes of the perovskites of up to 0.37 eV for MAPbI$_3$, 0.25 eV for MAPbBr$_3$ and CsPbBr$_3$ and 0.11 eV for the already large band gap MAPbCl$_3$. The magnitude of these shifts depends on the material, in line with the trend of the respective bandgap energies and the associated reduced exciton masses *(15, 41)*. Our methods are not limited to single-halide precursors. Color tuning via a combination of size effects and adjusting halide compositional stoichiometry is possible in both npSi and npAAO (fig. S1), eventually allowing accessing the whole visible spectral range. The uniform blue shifted emission of nanocrystalline perovskites in npAAO under UV illumination demonstrates that the whole sample area emits homogeneously over square centimeters (Fig. 1F). The blue shifts are even more pronounced for the npSi matrix: here we achieve red emission (640 nm) from NIR emitting bulk MAPbI$_3$ and blue color (480 nm) from the normally green bulk CsPbBr$_3$ (Fig. 1G).

**Color tunability**

Precise nanopore-size tunability is a compelling advantage of the npSi system. We use a galvanostatic method (*42*) that allows pore size tuning by changing the anodization current density (detailed in Materials and Methods). The as-prepared porous silicon samples are hydride terminated and weakly photoluminescent. We observed that a radio-frequency oxygen plasma treatment completely eliminates the PL of the porous silicon samples (fig. S2). This procedure accomplishes the dual role of extinguishing native luminescence and creating a surface that is highly wettable by the perovskite precursor solution. We find that the PL of $MAPbI_3$ is increasingly blue shifted as the pore size is reduced, from initially bulk-like NIR emission at 791 nm across the visible red range, with a maximal hypsochromic shift to 640 nm (Fig. 2A). $MAPbBr_3$ (Fig. 2B) and $CsPbBr_3$ (Fig. 2C) behave similarly; here the normally green-emitting materials can be shifted to give visibly cyan (485 nm) and blue (479 nm) emission. In all these cases, the peak emission wavelength varies linearly with the anodization current used to prepare the pores (fig. S3), indicating a dependence of band gap on nanopore size. Even the emission of the wide band-gap $MAPbCl_3$ can be shifted further into the UV range, well below 400 nm (Fig. 2D). Hypsochromic shifts obtained with transparent, thin film npAAO templates on glass slides (250-nm npAAO; see Materials and Methods) are also shown (Fig. 2, A to C). The observed shifts qualitatively agree well with reported blue shifts for different methods of obtaining nanocrystals or nanoplatelets *(20, 28)*. We find a PL quantum yield (PLQY) of up to 25 % ($\bar{x}$ = 13.76 %, σ = 9.17 %, n = 4) for $npMAPbI_3$, up to 60 % ($\bar{x}$ = 36.04 %, σ = 16.23 %, n = 4) for $ncMAPbBr_3$ and up to 90 % ($\bar{x}$ = 54.84 %, σ = 31.98 %, n = 4) for $ncCsPbBr_3$ in npAAO (see Materials and Methods). The PLQY of our nanoparticle emitters is significantly higher than what we observe for the bulk semiconductor films (<1 to 2%) and fits well within the range reported for other nanosized perovskite materials *(27, 28, 30, 43)*. Time resolved PL measurements (fig. S4) show a short lifetime for our nanocrystals (0.3 to 3 ns), whereas bulk films are longer-lived (3 to 25 ns). This behavior is known for the quantum dot systems *(15, 44)* and also observed for MAPbI3 films with varying crystallite dimensions *(45)* and for MAPbBr3 in mesoporous silica *(27)*.

Here, we here show that the confinement effects enable broad tunability windows in emission color from single-halide precursors, leaving the spectral regions from 640 to 530 nm and 479 to

405 nm unaddressed. To illustrate the generality of our approach, we complete these gaps by complementing the confinement effects with mixed halide stoichiometry. We demonstrate this with mixed Iodide-Bromide nanocrystals in npAAO resulting in yellow emission, and with mixed Bromide-Chloride perovskites in npSi resulting in deeper blue emission (fig. S1). Besides determining crystallite size, our thin film oxide templates enhance the stability and largely prevent degradation of the nanocrystals when excited with blue light (405 nm) under ambient conditions. Whereas bulk films decay rather rapidly under these conditions, the PL intensity of ncMAPbBr$_3$ in 250-nm-thick npAAO (Fig. 2E) initially increases by roughly 40 % and then stabilizes without further changes in shape of the PL signal (Fig. 2F). Similar results are obtained when comparing CsPbBr$_3$ with ncCsPbBr$_3$ (Fig. 2, G and H), wherein an initial increase is followed by delayed decay back to the original values. We believe the alumina matrix serves as encapsulation largely preventing ingress of oxygen and water. The initial increase in PL intensity may be attributed to light curing with oxygen, reducing non-radiative charge recombination by deactivating the respective trapping sites (*46*). Encapsulating ncMAPbBr$_3$ and ncCsPbBr$_3$ samples with epoxy and glass slides further corroborates these findings because the PL intensity remains essentially unchanged in the absence oxygen and water. This in turn demonstrates the stability of our nanocrystals under photoexcitation, without signs for light-induced degradation.

**Size analysis**

To investigate whether the observed hypsochromic shifts are indeed size-induced, we perform structural studies. Electron microscopy is used to test for the porosity and nucleation profile, as well as to get a first indication about particle sizes. We prepare focused ion beam (FIB)-milled lamellae of npAAO and find columnar structures on large areas using BF-STEM (fig. S5). From the uniform distribution of nanopores (top-view SEM, fig. S5), we estimate a template porosity of 15 %. Energy dispersive X-ray spectroscopy scans (EDX) confirm infiltration of the npAAO with perovskite nanocrystals (fig. S5). The EDX analysis also confirms that the bright regions (due to z-contrast) imaged on exfoliated npSi/ncMAPbI$_3$ flakes in high-angle annular dark field (HAADF, Fig. 3A) are perovskites. EDX line scans (Pb L line) from the thinnest parts of the flakes allow us to infer a size of ~4 nm for the perovskite crystals (Fig. 3B). In order to quantify the size of the perovskite crystallites that form in npSi, we use microfocus high-energy X-ray depth profiling in transmission geometry (fig. S6). Three npSi/ncMAPbI$_3$-samples (anodized at 15 mA cm$^{-2}$, 25 mA cm$^{-2}$ and 30 mA cm$^{-2}$) are probed in order to correlate anodization current density with resulting crystal size. The microfocus beam (FWHM, 2 - 5 µm; compare figs. S6 and S7) allows us to verify the formation of perovskite nanocrystals throughout the entire thickness of the npSi layers. The obtained diffraction patterns (figs. S6 and S7) are a superposition of the background due to the npSi and Debye rings of the perovskite crystallites within the pores. The background corrected azimuthally averaged MAPbI$_3$ scattering signal is extracted as a function of the scattering vector **q**. The maximal perovskite signal for each sample is shown in Fig. 3C [for intensity profiles at all depths see fig. S6 (30 mA cm$^{-2}$) and S7 (15 mA cm$^{-2}$ and 25 mA cm$^{-2}$)]. The observed peak positions obtained from the Debye rings are indexed to a tetragonal structure with the lattice constants a = b = 8.90(3) Å and c = 12.71(5) Å. These lattice constants match the literature values of Stoumpos *et al.* (*47*) a = b = 8.849(2) Å and c = 12.642(2) Å within 0.6 %. A complete peak indexing is given in fig. S6. All MAPbI$_3$ peaks show strong broadening due to finite crystal size. Using the Scherrer equation (*48*), we calculate the depth-dependent crystallite size (Fig. 3, D to F) from the FWHM of the diffraction peaks (see figs. S6 and S7 for details). The amount of perovskite material accumulating at a specific depth can be read off from the depth-dependent diffracted power, that is the integrated intensity of the perovskite diffraction peaks. In all cases, we find a rather narrow distribution of crystallite size versus depth, allowing us to average over the whole layer thickness using the amount of MAPbI$_3$ at the specific depth as weights. We obtain average MAPbI$_3$ crystallite sizes of 1.8 nm, 2.1 nm and 4.5 nm for 15 mA cm$^{-2}$ npSi, 25 mA cm$^{-2}$ npSi and 30 mA cm$^{-2}$ npSi respectively. This is the same order of magnitude as values reported for the exciton Bohr radius in MAPbI$_3$ (2.2 and 2.8 nm) *(49,50)*, accounting for the strongly blue-shifted emission observed in these nanocrystals. We further estimate the pore size and porosity of a 15 mA cm$^{-2}$ npSi scaffold from additional X-ray experiments (fig. S8). A film porosity of 57 % is determined from X-ray transmission measurements. Small angle X-ray scattering (SAXS) experiments following the method of Porod *(51)* reveal an average pore diameter of 4.0 nm. Comparing this with the crystal size analysis, we find that the formed perovskite nanocrystals are smaller than the surrounding pores (1.8 nm vs. 4.0 nm). This indicates that the pores act as weakly connected nanoreactors, placing an upper bound on the perovskite nanocrystal size. Because of this enhancement of confinement, nanocrystals consisting of only a few unit cells (around 6 in the case of 15 mA cm$^{-2}$ npSi) can be achieved that result in pronounced PL blue shifts. Our structural characterization reveals that the average crystallite size decreases with decreasing pore size. This directly translates into an increase of the PL peak emission energy (Fig. 3G), altogether providing strong evidence that the observed PL blue shifts are indeed due to quantum size effects.

**Light-emitting diodes**

Incorporating nanoscale (less than 10 nm) perovskite emitters as active layers in light emitting diodes is currently pursued with colloidal dots and platelets, promising low-cost high color purity photonic sources (*9–11*, *52–55*). The possibility and ease of three-dimensional structuring incorporated in thin film device architectures are unique benefits of our nanoporous solid templates. Translating this concept of solid-state confinement into devices requires electrical addressability of the nanotemplates. We demonstrate that insulating npAAO filled with perovskite aids the formation of conductive nanostructures, enabling low-voltage electroluminescent diodes (Fig. 4A) with narrow and blue shifted emission. In order to fabricate nanoporous perovskite nanocrystal light-emitting diodes (npPeLEDs), a bilayer of Ti and Al is deposited onto transparent fluorine-doped tin oxide (FTO). Aluminum is anodized by electrochemical oxidation to obtain npAAO (60 nm film, Fig. 4B and fig. S8) and then infused with precursor solution to yield perovskite nanocrystals. We observe uniform PL over large areas after the npAAO is filled with perovskites from solution (Fig. 4C), indicating the suitability of the titania/npAAO/perovskite composite films for diodes. Titanium is not porosified by the electrolyte during anodization, forming compact $TiO_2$ that acts as electron injection layer in the sandwich structure. The device is completed by a hole conducting polymer (HTL) top layer and MoOx/Ag contacts (Fig. 4A). Devices prepared with ncMAPbI$_3$ give red electroluminescence (EL) peaking at 731 nm (Fig. 4D), i.e. ~60 nm blue shifted from the NIR emitting bulk. For CsPbBr$_3$ devices, we find a sharp green-cyan EL centered at 518 nm. Its FWHM of 17 nm (Fig. 4E) is in line with the results of Li *et al*. *(53)* shown for CsPbBr$_3$ nanocrystal devices. These diodes turn on at ~2.5 V and give around 300 cd m$^{-2}$ operating at 5 V (Fig. 4F). At this voltage, the J-V characteristics show that npAAO/perovskite layers carry a current density of around 420 mA cm$^{-2}$. Our perovskite LEDs have a current efficiency of ~0.03%; however, the demonstrated PLQY of up to 90% for our perovskite nanocrystals emitters suggests that higher efficiencies are achievable through further device optimization. This is expected to match the performance of current state-of-the-art 2D- and 3D-based perovskite LEDs *(10, 13, 43, 52, 53, 55)* while preserving the ease of color tuning. Control samples prepared without perovskite were electrically insulating. The optoelectronic quality of the composite nanoporous layer is exemplified by uniform EL over continuous areas as large as 15 mm$^2$ (Fig. 4G).

**Discussion**

We demonstrated blue emission from ncCsPbBr$_3$ in npSi, and cyan electroluminescence in npAAO based LEDs. Strong blue shifts are observed for the silicon-based templates, as their small pore sizes allow for crystallite sizes as small as 2 nm. Future electrical addressing of these templates seems feasible through down-scaling of the npSi matrix to sub-100-nm thicknesses via advanced silicon patterning or porosification protocols. Reducing the pore size in npAAO is expected to result in more pronounced blue shifts, feasible routes are optimization of the anodization protocol or narrowing the pore diameter by deposition of conformal interlayers, for example via atomic layer deposition. The electrical performance of our npPeLEDs is expected to further improve by reducing non-radiative recombination within the semiconductor and at the TiO$_2$/perovskite and perovskite/HTL interfaces. The npPeLEDs presented herein already reach indoor-display brightness levels, demonstrating the potential of solid-state confinement for perovskite photonic sources. Color tuning in single-halide perovskites through quantum size effects may alleviate stability issues (*56*). Forming ligand-capped 2D or 3D nanocrystals directly within our nanoporous solids may open additional routes toward size and shape control in integrated, device compatible thin film structures. Future work should focus on fundamental research, exploring complex lateral and depth-resolved nanostructuring for manipulation of the optoelectronic properties. Potential applications range from photon detectors and (polarized) electroluminescence devices to single photon sources and metasurfaces.

**Materials and Methods**

**Perovskite precursor solutions**

All chemicals and solvents were purchased from commercial suppliers and used as received, if not stated otherwise. Methylammonium bromide (MABr$_3$) was synthesized from methylamine [33 weight % (wt%) in absolute ethanol, Sigma Aldrich] and hydrobromic acid (48 wt%, aqueous, Sigma Aldrich), and purified utilizing diethylether (VWR) and absolute ethanol (Merck) as described in literature (*57, 58*). Methylammonium chloride (MACl$_3$) was prepared via an analogous protocol using hydrochloric acid (37 wt% aqueous, Merck). Lead bromide (PbBr$_2$, 99.999%), dimethylsulfoxide (DMSO; ≥ 99%) and methylammonium iodide (MAI) were supplied by Sigma Aldrich, Merck and Dyesol, respectively. Lead iodide (PbI$_2$, 99.9985 %), cesium bromide (CsBr, 99.999%) and dimethyl formamide (DMF, 99.8%) were purchased from Alfa Aesar. All parent solutions were passed through polytetrafluoroethylene (PTFE) syringe filters (0.45 µm; Whatman). MAPbI$_3$: PbI$_2$ (922 mg, 2.00 mmol), H$_3$CNH$_3$I (334 mg, 2.10 mmol) and DMF (2.25 mL) were mixed to yield a clear yellow solution. MAPbBr$_3$: PbBr$_2$ (368 mg, 1.00 mmol), H$_3$CNH$_3$Br (124 mg, 1.11 mmol) and DMF (1.10 mL) were heated to 50 °C overnight to yield a clear colorless solution. MAPbCl$_3$: PbCl$_2$ (280 mg, 1.01 mmol), H$_3$CNH$_3$Cl (76 mg, 1.13 mmol) and DMSO (500 µL) were heated to 60 °C to yield a clear colorless solution. CsPbBr$_2$: Stirring a mixture of PbBr$_2$ (367 mg, 1.00 mmol), CsBr (215 mg, 1.01 mmol) and DMSO (2.4 mL) at 70 °C overnight resulted in a clear colorless solution. Iodide-bromide and bromide-chloride containing precursors were prepared from single halide solutions by mixing in a 80:20 volume ratio.

**Nanoporous silicon (npSi)**

Two types of boron-doped p-type silicon wafers with thickness of 0.5 mm and <100> orientation were used, either "low-doped" $R_0$ = 8 to 20 ohm cm, or "highly-doped" $R_0$ = 0.01 ohm cm. The wafers were one-side polished, with an alkaline-etched backside. The Si wafers were cut into 1"×1" squares, cleaned by ultrasonication sequentially in 2% solution of chemical detergent Hellmanex III, deionized water, acetone and 2-propanol, followed by the RCA standard cleaning steps (*59*). Next 250 nm aluminum contacts were evaporated onto the alkaline-etched side. After deposition of the Al, the samples were annealed in N$_2$ atmosphere at 450 °C for 20 min to ensure an Ohmic Al/p-Si contact. The polished surface of the substrate was nanostructured by anodization in the Standard Etch Cell as defined by Sailor (*42*). Contact to the silicon substrate was established by directly contacting aluminum foil with the aluminized back of the substrate, while the counter electrode was a platinum coil. Anodization was conducted in galvanostatic mode using a Keithley 2400 Sourcemeter with variable current density: 5-30 mA cm$^{-2}$ for low-doped Si, and 150-370 mA cm$^{-2}$ for highly-doped Si. Anodization was performed for 20 min for all the samples. The electrolyte consisted of a mixture of 48% HF and ethanol in volumetric ratio of 1 to 1 for low-doped Si and 4 to 1 for highly-doped Si. Following anodization, the electrolyte was removed with a plastic micropipette, the sample was thoroughly washed with ethanol and dried under an N$_2$ stream. Next the npSi samples were treated with oxygen plasma (Plasma Etch PE-25-JW) at 50 W for 5 min, resulting in a highly hydrophilic surface that is readily wetted by the DMF and DMSO perovskite precursor solutions. The solutions were deposited by spin coating at 2400 rpm for 7 s and afterwards annealed at 115 °C for 30 min in ambient atmosphere.

**Anodic aluminum oxide nanotubes (npAAO)**

1×1" glass substrates 1 mm thick were cleaned by ultrasonication sequentially in Hellmanex III chemical detergent solution (aqueous, 2 volume %), deionized water, acetone and 2-propanol, followed by oxygen plasma cleaning for 5 min at 50 W. Varying thicknesses (20 nm - 1µm) of aluminum were thermally evaporated onto the glass slides at a rate of 1-5 nm s$^{-1}$ and a base pressure of ~1×10$^{-6}$ mbar. The samples were then anodized potentiostatically using a Keithley 2400 Sourcemeter in an aqueous 0.2 M oxalic acid solution and a platinum foil counter electrode. Small pores (< 8 nm) were achieved using an anodization voltage of 5 V. The aluminum films were contacted using an alligator clip and partially immersed into the electrolyte. The region at the air-electrolyte interface was masked with polyimide tape to prevent rapid electrochemical etching at the top of the sample. Anodization was continued until the steady-state anodization current rapidly dropped, signalling that the conductive aluminum layer had been completely consumed. At this point the npAAO film was visibly transparent and featured a blue-green iridescence. The npAAO samples were then rinsed with 18 MΩ water, dried, and treated with oxygen plasma for 5 min at 50 W prior to infiltration with perovskite precursor solutions.

**Nanoporous perovskite nanocrystal LEDs**

Patterned FTO coated glass slides (Xin Yan Technology Ltd., 1"×1" with a centred, 1.35 cm wide and 2.5 cm long stripe of FTO with a sheet resistance of 15 ohms per square) were polished with a titanium and silicon oxide containing polishing paste to improve surface qualities. The FTO coated glass slides were further cleaned using sequential ultrasonication as described for the npAAO preparation and subsequently treated with oxygen plasma (50 W for 5 min). Layers of titanium (15 nm) and aluminum (40 nm) were evaporated sequentially on top of the FTO in the same thermal evaporation system without breaking vacuum with deposition rates of 0.1 and 1 nm s$^{-1}$ at a base pressure of ~1×10$^{-6}$ mbar. Anodization of the double layer was performed as described above. This one-pot anodization protocol resulted in an optically transparent nanoporous alumina scaffold atop a compact titania film. This functional oxide double layer architecture served both as electron injection contact, and as insulating nanoscale template for the formation of conjoined perovskite nanocrystals. After rinsing with 18 MΩ water and drying, samples were treated with oxygen plasma (50 W for 5 min). The perovskite precursor solutions (parent solution of CsPbBr$_3$ 1:22 diluted with DMF or parent solution of MAPbI$_3$ 1:15 diluted with DMF) were deposited by spin-coating at 2400 rpm for 7 s. These recipes allowed filling of the 60-nm-high porous matrix without the formation of bulk perovskite layers, resulting in electrically addressable conjoined nanocrystals. The samples were annealed at 115 °C for 30 min in ambient atmosphere. Poly[9,9-dioctylfluorenyl-2,7-diyl] end capped with N,N-bis(4-methylphenyl)-aniline (F8, purchased from ADS, 0.5 wt%), dissolved in chlorobenzene, was spin-coated at 3000 rpm until dry yielding a hole-transporting polymer layer. Molybdenum(VI) oxide (20 nm) and silver (100 nm), thermally deposited at rates of 0.03 and 1 nm s$^{-1}$ through a metal shadow mask at a base pressure of ~1×10$^{-6}$ mbar completed the device.

**High energy X-ray diffraction**

The measurements were performed at the high energy beamline (P07) at PETRA III at DESY (Deutsches Elektronen-Synchrotron). The 30 mA cm$^{-2}$ npSi/ncMAPbI$_3$ sample was measured with an X-ray beam with the energy of 98.5 keV, which was focused down to a spot of 5 × 50 µm² (vertical × horizontal FWHM) at sample position using Al compound refractive lenses. For all other samples the respective values were 80.0 keV and 2 x 30 µm². The diffracted intensity was recorded 2000 mm behind the sample by a Perkin Elmer XRD 1621 Flat Panel detector. In order to distinguish clearly between the signal of the perovskite and background resulting from the substrate, a reference npSi sample without perovskite was measured for each beam configuration. For data processing, Igor Pro (WaveMetrics) including the "Nika" package (*60*) and OriginPro (OriginLab) were used (see Supplementary Materials for details).

**Small angle X-ray diffraction**

The samples were measured in transmission with the surface perpendicular to a 20.0 keV X-ray beam at beamline P08 at PETRA III at DESY. The scattered intensity was detected by a Perkin Elmer XRD 1621 Flat Panel detector at a distance of 2440 mm behind the sample. For data analysis the "Nika" package (*60*) for Igor Pro (WaveMetrics) and SASfit *(61)* were used.

**Scanning Transmission Electron Microscopy**

The npSi specimens were prepared by mechanical exfoliation from the porous areas of the sample on a holey carbon TEM grid. The npAAO specimens were prepared by focused ion beam milling (1540 CrossBeam Scanning Electron Microscope, ZEISS). A thin gold layer was sputtered onto the specimen prior to milling. TEM lamellas were prepared by standard FIB cutting, lift out and transfer to a TEM grid, followed by a final thinning to about 100 nm with an ion energy of 5 keV to minimize preparation artefacts. Scanning microscopy was performed in bright filed (BF) and high angle annular dark field (HAADF) modes, sensitive to the z-contrast of the Pb-rich regions. Line and mapping analyses where carried out using EDX. All specimens were investigated with a

JEOL JEM-2200FS transmission electron microscope with STEM mode operated at 200 kV, and an Oxford SDD X-maxN (80 mm²) EDX-system.

**Sample characterization**

Surface SEM measurements were made using the ZEISS 1540XB CrossBeam SEM. Optical microscopy images were recorded using a Nikon Eclipse LV100ND microscope with up to ×400 magnification and a photoluminescence filter cube accessory. PL spectra were recorded on a photomultiplier tube-equipped double-grating input and output fluorometer (PTI). Electroluminescence measurements were performed using a Shamrock SR-303i monochromator, an Andor iDus Si-CCD camera and a Keithley 2400 Sourcemeter. The optical power output of the npPeLEDs was measured with an Agilent B1500 Parameter analyser and a calibrated silicon diode (Hamamatsu S2281).

**PL Quantum Efficiency**

Samples were prepared as described before using 150 nm thick layer of npAAO on glass. Several samples were prepared for each of the systems with varying concentration of precursor solutions in range between 0.06 – 0.1 M. The PLQE was measured using a method previously described in literature (*62*). The measurement was performed inside an integrating sphere in a fluorometer (Photon Technology International) and the excitation wavelength was 405 nm with approximately 0.4 mW cm$^{-2}$ illumination intensity.

**PL stability measurements**

Samples were prepared with 150 nm of npAAO on glass and 0.08 M solution of $CsPbBr_3$ or $MABr_3$. Samples were placed into fluorometer (Photon Technology International) and spectra were recorded for about 28 hours. Peak intensity was used for the data evaluation. Peak wavelength was monitored and showed no variation throughout the duration of the measurement.

Encapsulated samples of perovskite nanocrystals in npAAO were sealed with UV curable epoxy (Ozilla E131) and glass sealed.

**Time-resolved PL lifetime measurements**

Bulk MAPbI3, MAPbBr3, and CsPbBr3 films were measured using an yttrium-aluminum-garnet–Nd laser (Spitlight Compact 100) emitting at 355 or 532 nm with a pulse length of ~10 ns and an energy of 50 mJ cm$^{-2}$ in both cases. The spot size was 5 mm in diameter. Signal detection was performed with a Shamrock spectrometer (SR-303i-A), equipped with an intensified charge-coupled device camera [Andor iStar DH320T-18U-73 (gate step, 2.5 ns; gate width, 2.5 ns)]. The samples were kept under vacuum during the measurement. Corresponding shorter-lived perovskite nanocrystals in 5–mA cm−2 npSi templates were investigated using a superconducting single photon detector (SSPD; Scontel Superconducting Nanotechnology) together with a timecorrelated single-photon counting system (PicoHarp 300 by PicoQuant). The time resolution of the SSPD system is about 300 ps. Samples were excited by a pulsed diode laser (405 nm) delivering ~1-ns pulses (FWHM) at repetition rates of 1 MHz and pulse energies of approximately 5 pJ. A microscope objective was used to focus the laser onto a spot with a 10-mm diameter on the sample surface. The sample emission was collected using the same microscope objective, and the photons were guided through a single-mode fiber to the SSPD. A long-pass filter was applied to block the excitation before entering the fiber connected to the SSPD.

**Supplementary materials**

Supplementary material for this article is available at http://advances.sciencemag.org/cgi/content/full/3/8/e1700738/DC1

fig. S1. Bandgap tuning via size effects and halide compositional stoichiometry in npSi and npAAO.

fig. S2. PL of native npSi and MAPbI3 nanocrystals in npSi.

fig. S3. Fine-tuning of the bandgap in npSi via variation of pore size.

fig. S4. Time-resolved PL lifetime.

fig. S5. Structural characterization of MAPbBr$_3$ nanocrystals in npAAO.

fig. S6. Wide-angle x-ray scattering (WAXS) study of MAPbI$_3$ nanocrystals in 30–mA cm$^{-2}$ npSi: Experiment and data analysis.

fig. S7. WAXS study of MAPbI3 nanocrystals in 15–mA cm$^{-2}$ npSi and 25–mA cm$^{-2}$ npSi.

fig. S8. Pore size estimation by Porod analysis of SAXS data of 15–mA cm$^{-2}$ npSi.

fig. S9. Surface SEM of npAAO on compact TiO$_2$. References *(63–67)*.

## Acknowledgments

We thank Florian Hackl for technical assistance with PL measurements.

## Funding

This work was supported by the European Research Council Advanced Investigators Grant "Soft-Map" for S.B. and the FWF Wittgenstein Award (Solare Energie Umwandlung Z222-N19) for N.S.S. H.G. acknowledges funding from the Austrian Federal Ministry of Science, Research and Economy and the National Foundation for Research, Technology and Development. J.M.R. and B.N. acknowledge funding by the Bavarian State Ministry of Science, Research, and Arts through the grant "Solar Technologies Go Hybrid (SolTech)" and by the DFG through SFB 1032. E.D.G. acknowledges support from the Knut and Alice Wallenberg Foundation within the framework of the Wallenberg Centre for Molecular Medicine at Linköping University. Parts of this research were carried out at the light source PETRA III at DESY (Deutsches Elektronen-Synchrotron), a member of the Helmholtz Association (HGF), using beamline P07 and P08.

## Author contributions

E.D.G. and M.K. conceived and supervised the research. S.D., E.D.G., and M.K. fabricated and characterized the samples and devices. C.U. prepared the perovskite solutions. S.D., H.H., and M.C.S. measured the EL and PL. J.M.R., A.B., U.R., F.B., and B.N. performed the x-ray experiments, and J.M.R., A.B., and B.N. analyzed the data. H.G. and G.H. performed the STEM/EDX measurements and analyzed the data. D.A. obtained the SEM images. S.D., J.M.R., S.B., E.D.G., and M.K. analyzed the data, designed the figures, and wrote the manuscript with comments from all co-authors. N.S.S., S.B., E.D.G., and M.K. coordinated the project.

**Competing interests**

The authors declare that they have no competing interests. Data and materials availability: All data needed to evaluate the conclusions in the paper are present in the paper and/or the Supplementary Materials. Additional data related to this paper may be requested from the authors.

**Citation**

S. Demchyshyn, J. M. Roemer, H. Groiß, H. Heilbrunner, C. Ulbricht, D. Apaydin, A. Böhm, U. Rütt, F. Bertram, G. Hesser, M. C. Scharber, N. S. Sariciftci, B. Nickel, S. Bauer, E. D. Głowacki, M. Kaltenbrunner, Confining metal-halide perovskites in nanoporous thin films. Sci. Adv. 3, e1700738 (2017).

# Figures

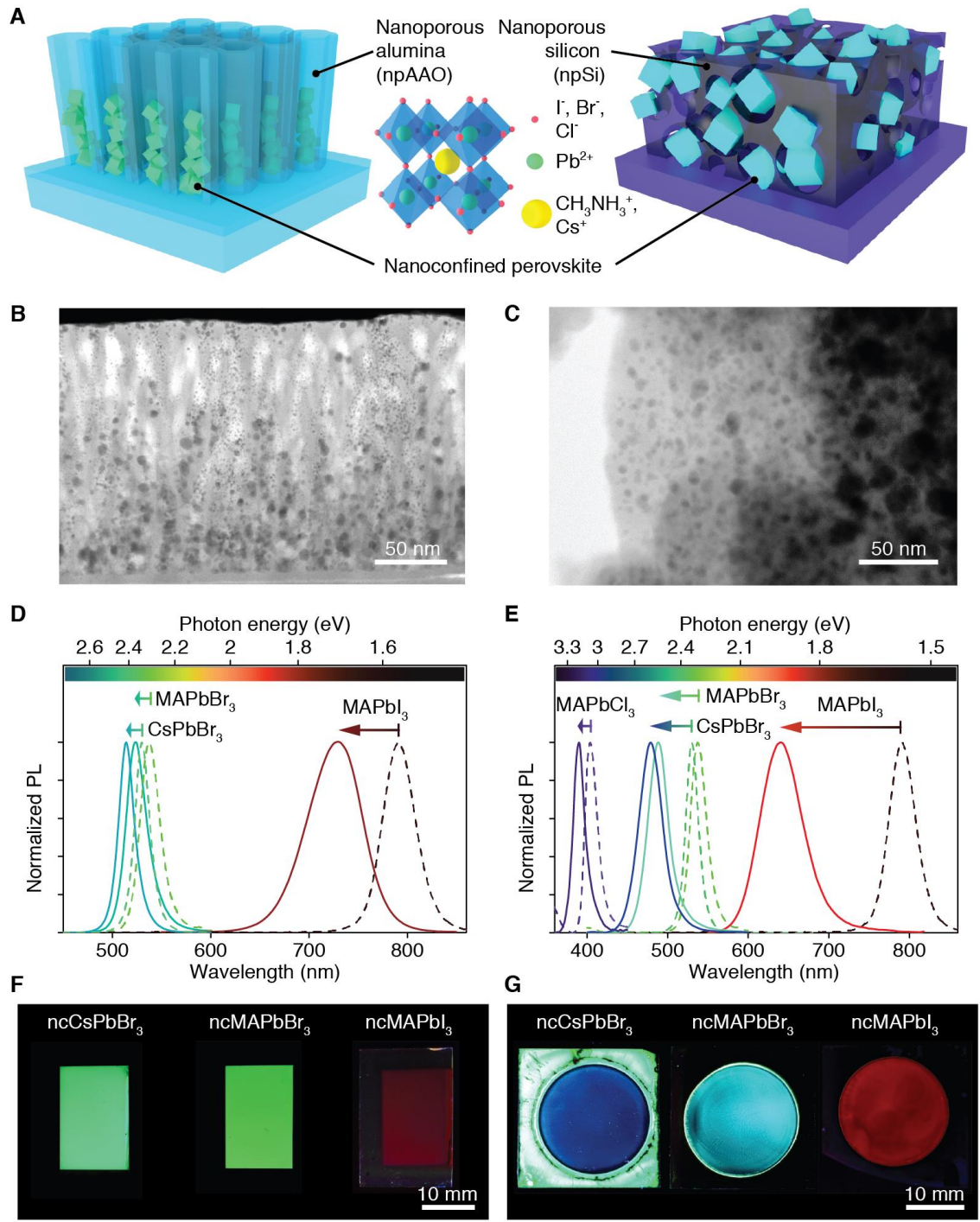

**Fig. 1. Metal-halide perovskites confined in nanoporous films.** (**A**) Schematic of nanoscale solid film templates (npAAO on the left and npSi on the right) infused with perovskite

nanocrystals. The chemical structure of methylammonium lead trihalide (ncMAPbX$_3$) and cesium lead trihalide perovskites (ncCsPbX$_3$), with chloride, bromide, and iodide (X = Cl, Br, I), is shown in the middle. **(B)** BF-STEM image of a 170-nm-high, ~100-nm-thick npAAO filament, indicating alumina nanopores of 6 to 8 nm in diameter, partially filled with conjoined perovskite nanocrystals. **(C)** Exfoliated flake of npSi filled with MAPbI$_3$, imaged by BF-STEM. Perovskite nanocrystals appear dark because of their diffraction and mass contrast. Nanocrystals on the right appear aggregated; this is an imaging artefact caused by an increase in flake thickness. **(D)** PL of ncMAPbI$_3$ grown in npAAO (solid red line), blue-shifted by 62 nm. ncMAPbBr$_3$ (solid green line) and ncCsPbBr$_3$ (solid cyan line) are shifted by 14 and 16 nm, respectively. Dashed lines show bulk film PL. **(E)** PL of perovskite-infiltrated npSi. These smaller pores result in a 150-nm shift for ncMAPbI3 (solid red line), 52 nm for ncMAPbBr$_3$ (solid cyan line), 51 nm for ncCsPbBr$_3$ (solid blue line), and 14 nm for ncMAPbCl$_3$ (solid purple line). Dashed lines indicate bulk film PL. **(F and G)** Photographs of square centimeter–scaled thin films of nanocrystalline perovskites under UV illumination: npAAO on glass slides **(F)** and npSi on Si wafers **(G)**. The circular areas in **(G)** are nanoporous.

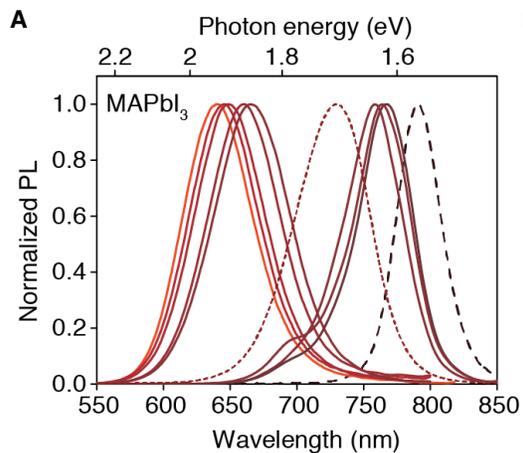
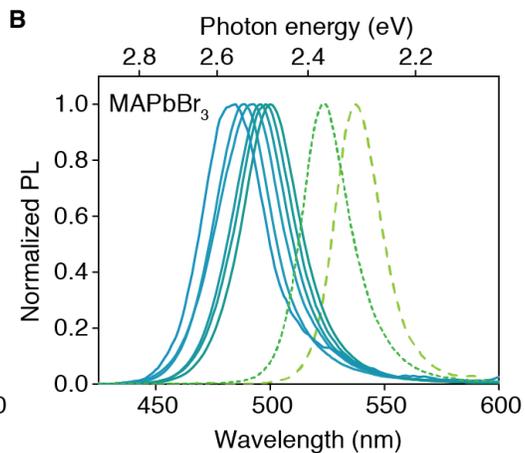
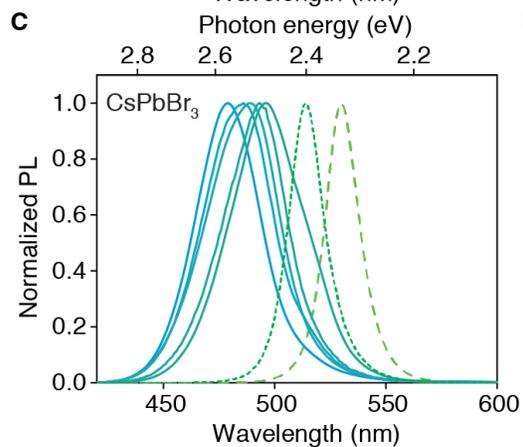
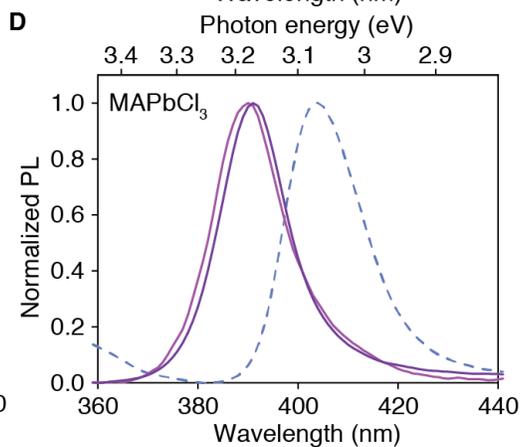
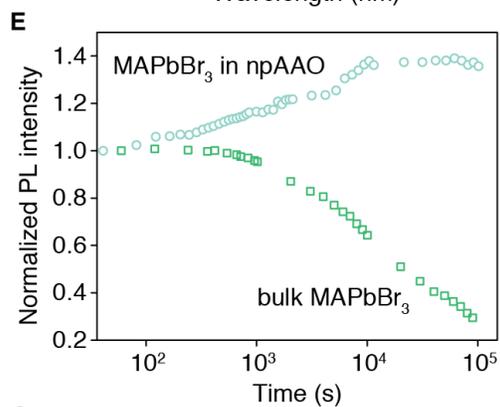
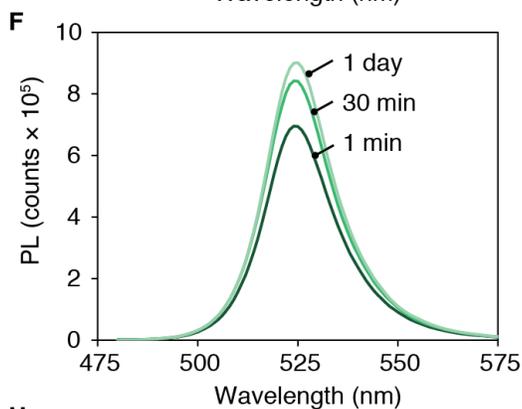
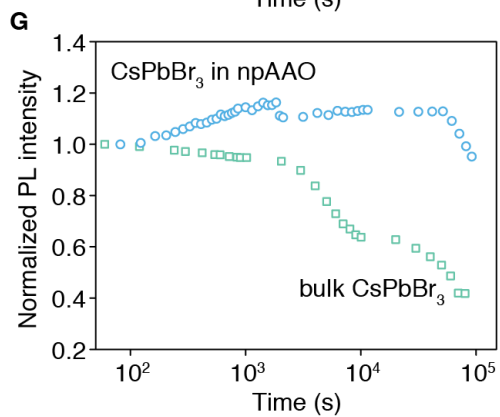
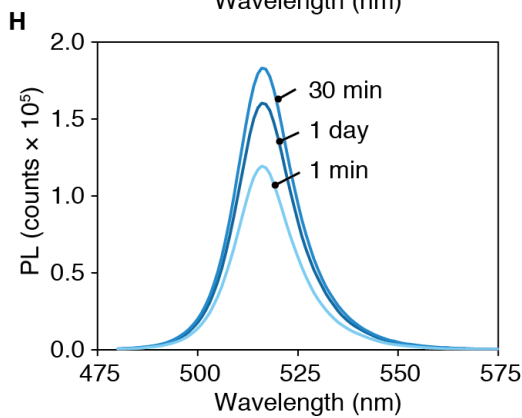

**Fig. 2. PL tuning and stability of perovskites confined in nanoporous films.** The dashed lines represent the bulk PL, whereas the dotted lines are the PL of perovskites in npAAO. Solid lines correspond to the PL of perovskites grown in npSi, with successive blue-shifted peaks originating from samples with progressively smaller pore sizes. (**A**) $MAPbI_3$ bulk versus crystals confined in npAAO and in eight differently sized npSi scaffolds. A clear transition from NIR emission to visible red is observed. (**B**) $MAPbBr_3$ bulk versus crystals confined in npAAO and six differently sized npSi scaffolds, showing emission shifting from green to blue. (**C**) $CsPbBr_3$ bulk versus confined nanocrystals in the same set of npAAO and npSi. (**D**) The UV-emitting $MAPbCl_3$ bulk compared with confined crystals in two of the smallest npSi matrices. Even this already wide-bandgap material can be blue-shifted via spatial confinement to emit below 400 nm. (**E**) Time evolution of the PL intensity during illumination at 405 nm under ambient conditions for $ncMAPbBr_3$ in npAAO (cyan empty circles) compared to bulk planar films of $MAPbBr_3$ on glass (green empty squares), evidencing encapsulation-like effects of the npAAO thin film. Sealing $ncMAPbBr_3$ samples with epoxy and glass slides (dark blue empty triangles) results in an essentially constant PL signal, demonstrating stability under continuous illumination without light-induced degradation. (**F**) PL signal for $ncMAPbBr_3$ in npAAO after 1-min (dark green), 30-min (green), and 1-day (light green) illumination at 405 nm. (**G**) Time evolution of the PL intensity during illumination at 405 nm under ambient conditions for $ncCsPbBr_3$ in npAAO (blue empty circles) and bulk $CsPbBr_3$ on glass (cyan empty squares). Epoxy- and glassencapsulated $ncCsPbBr_3$ samples (dark blue empty triangles) again show stable PL and no photodegradation. (**H**) PL signal for $ncCsPbBr_3$ in npAAO after 1-min (light blue), 30-min (blue), and 1-day (dark blue) illumination at 405 nm.

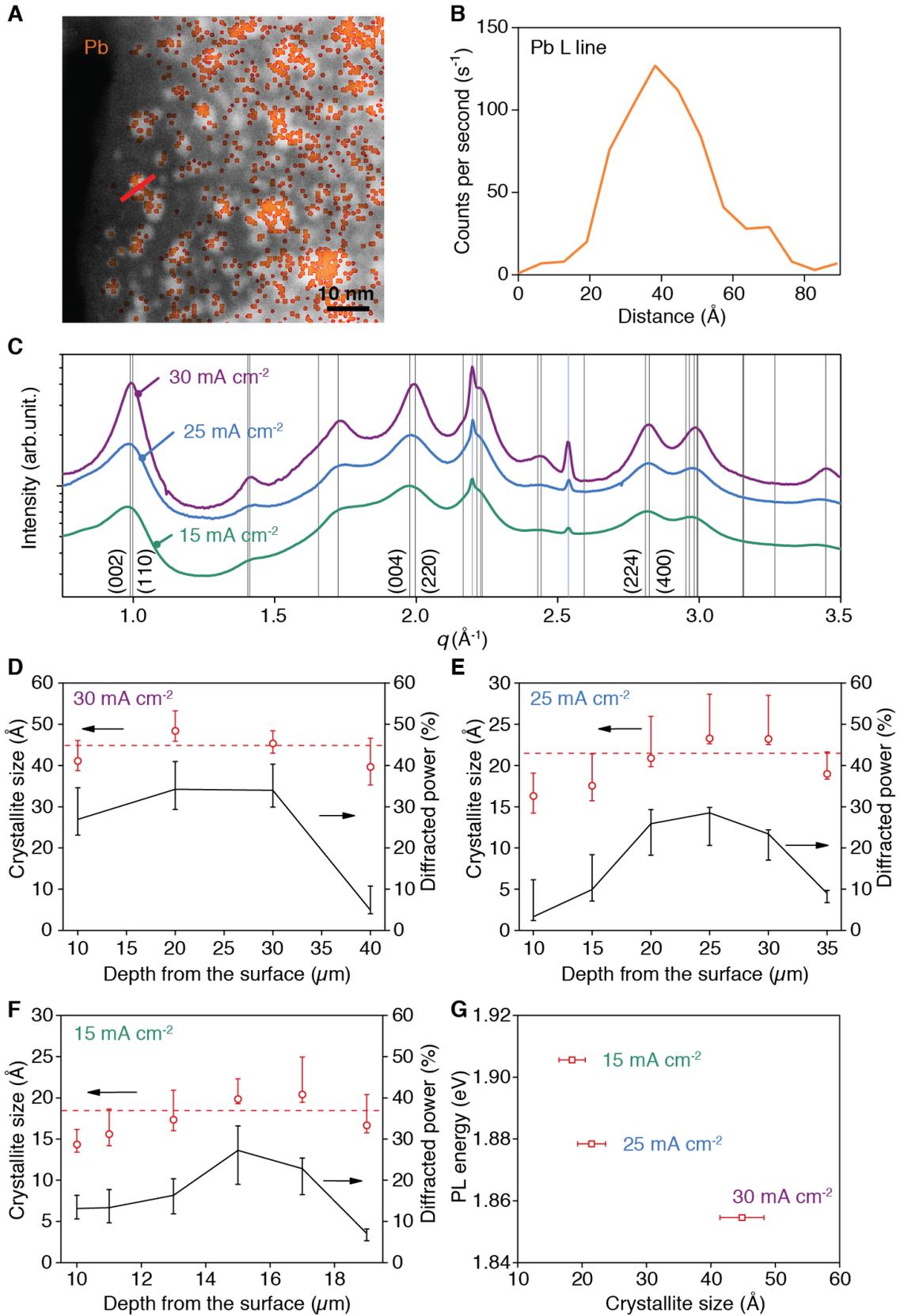

**Fig. 3. Depth-resolved structural characterization of perovskite nanocrystals in npSi films.**
**(A)** HAADF-STEM image of an exfoliated npSi/ncMAPbI$_3$ flake. The bright regions correspond to perovskite (due to z-contrast), as confirmed by EDX analysis. The Pb signal is plotted as orange overlay. **(B)** EDX line scan (Pb L line) of a single crystallite from the thinnest part of the flake [indicated as red line in (A)]. We find a crystallite size of 4 nm for 30–mA cm$^{-2}$ npSi. **(C)** Background corrected, azimuthally averaged x-ray diffraction profiles of MAPbI$_3$ in 30–mA cm$^{-2}$ npSi (violet line), in 25–mA cm$^{-2}$ npSi (blue line), and in 15–mA cm$^{-2}$ npSi (green line) as a function of the scattering vector **q**. The curves are normalized and vertically shifted for clarity. The log-scale highlights the increasing broadening of all MAPbI$_3$ peaks with decreasing current density due to crystallite size reduction, suggesting increasingly strong confinement. The peaks are indexed to a tetragonal structure with the lattice constants a = b = 8.90(3) Å and c = 12.71(5) Å. **(D to F)** Crystallite size and diffracted power of the MAPbI$_3$ signal as function of depth for **(D)** 30–mA cm$^{-2}$ npSi/ ncMAPbI$_3$, **(E)** 25–mA cm$^{-2}$ npSi/ncMAPbI$_3$, and **(F)** 15–mA cm$^{-2}$ npSi/ncMAPbI$_3$. Red circles: Depth-dependent crystallite size. Black line: Integrated intensity of the perovskite diffraction signal as a measure for the amount of perovskite at a specific depth. Dashed red line: Weighted average of the crystallite size. Error bars correspond to the 1-σ values of the uncertainty resulting from data analysis and the initial resolution of the experiment. **(G)** PL peak emission energy against the average size of crystallites formed in three npSi layers prepared with indicated anodization current density. Error bars correspond to the SD of the crystallite size.

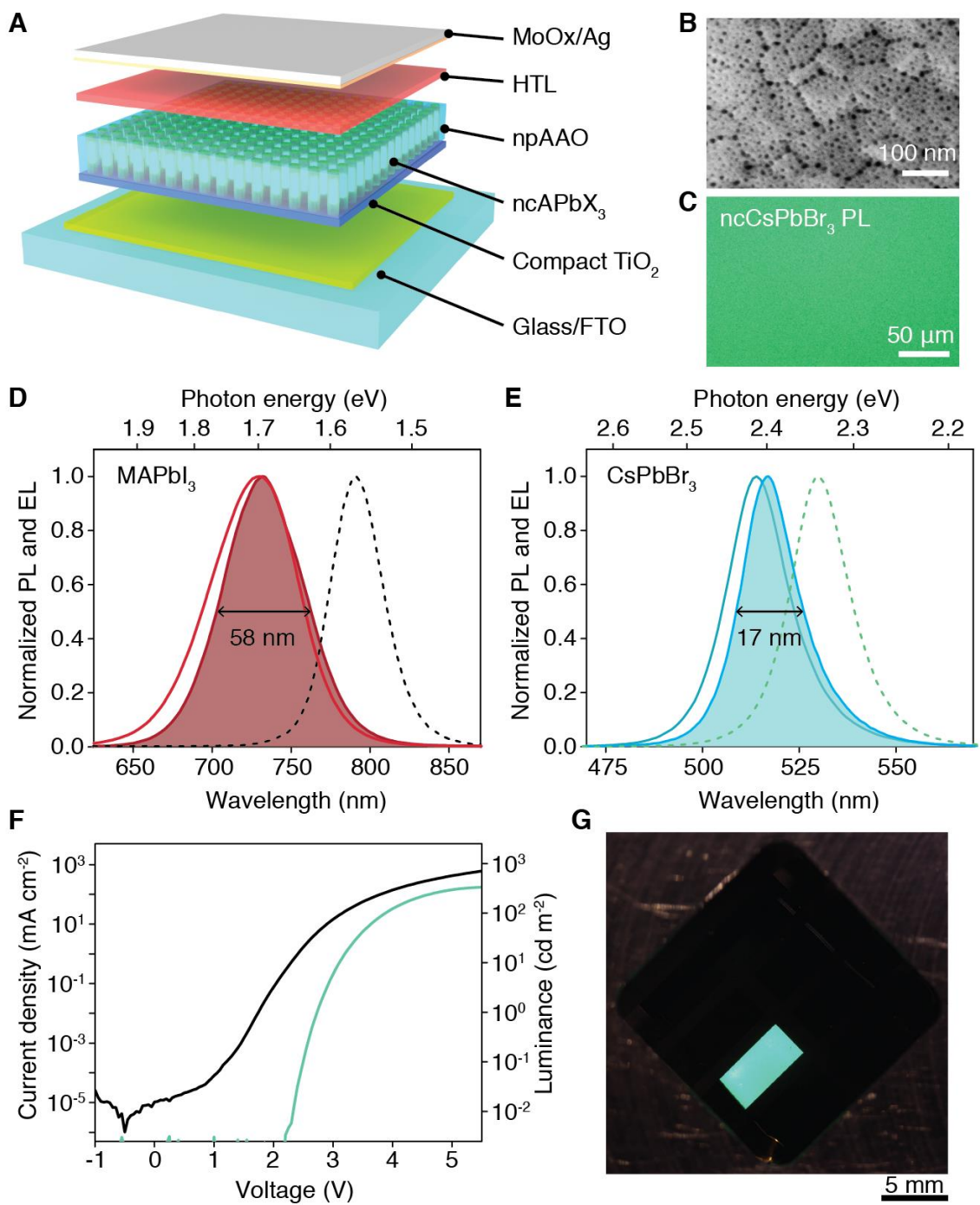

**Fig. 4. Perovskite-infused nanoporous thin films for LEDs. (A)** Schematic of npPeLEDs. Devices are fabricated on glass/FTO by anodizing a titanium/aluminum bilayer. This electrochemical procedure produces npAAO with compact TiO$_2$ below. Precursor solution infused to the top of the pores yields conducting and electroluminescent nanocrystalline

perovskite domains. A polymeric HTL and hole-injecting contact of MoOx/Ag complete the device stack. **(B)** SEM of anodized Ti/Al, showing the npAAO layer with pores of ~6-nm diameter on top of the scale-like Ti/TiO$_2$ domains below. **(C)** PL micrograph image taken of a Ti/npAAO layer infused with CsPbBr$_3$, evidencing uniform light emission across large areas of the sample. Excitation filter, 465 to 495 nm; barrier filter, 515 to 555 nm. (D) EL of an ncMAPbI3 diode, (filled curve) centered around 731 nm, with bulk PL emission (dashed line) and PL of the ncMAPbI$_3$ diode (solid curve) plotted for comparison. **(E)** EL (filled curve) of an ncCsPbBr$_3$ diode with PL emission of the bulk (dashed line) and the ncCsPbBr$_3$ diode is shown (solid line). Cyan-green narrowband EL peaking at 518 nm is observed. **(F)** J-V characteristic of an ncCsPbBr$_3$ diode (black trace) plotted together with luminance (cyan). Devices turn on at ~2.5 V with a luminance of ~330 cd m$^{-2}$ at 5 V. **(G)** Photograph of a CsPbBr$_3$ diode while operating at 4 V, displaying cyan-green EL over the entire area of the pixel (15 mm$^2$).

Supplementary material for

# Confining Metal-Halide Perovskites in Nanoporous Thin Films

Stepan Demchyshyn, Janina Melanie Roemer, Heiko Groiß, Herwig Heilbrunner, Christoph Ulbricht, Dogukan Apaydin, Anton Böhm, Uta Rütt, Florian Bertram, Günter Hesser, Markus Clark Scharber, Niyazi Serdar Sariciftci, Bert Nickel, Siegfried Bauer, Eric Daniel Głowacki, Martin Kaltenbrunner

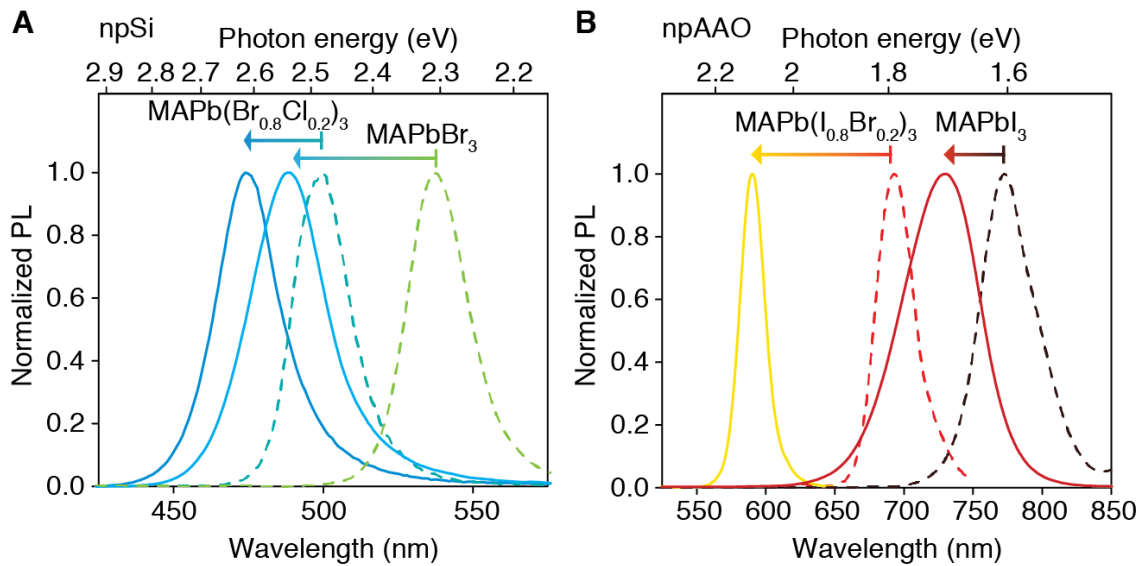

**fig. S1. Bandgap tuning via size effects and halide compositional stoichiometry in npSi and npAAO.** **(A)** Infusing npSi (anodized at a current density of 10 mA cm$^{-2}$) with a mixed BromideChloride (80:20) precursor solution results in the growth of perovskite nanocrystals containing both halides, exhibiting blue photoluminescence centered at 474 nm (solid blue trace). We observe a 26 nm blue-shift compared to bulk films of MAPb(Br$_{0.8}$Cl$_{0.2}$)$_3$ (dashed cyan trace) due to size effects. PL of single-halide MAPbBr$_3$ bulk films (dashed green trace) and nanocrystals grown in a 10–mA cm$^{-2}$ template (solid cyan trace) are shown for comparison. **(B)** Confining nanocrystals formed from a mixed Iodide-Bromide (80:20) precursor solution within npAAO templates results in yellow photoluminescence (solid yellow trace) centered at 590 nm,

blueshifted by 103 nm from the red-emitting bulk MAPb(I$_{0.8}$Br$_{0.2}$)$_3$ films (dashed red line). Emission from single-halide MAPbI$_3$ bulk film (dashed black line) and from nanocrystals in an identical npAAO matrix (solid deep red line) are also shown. These results further corroborate the universality of our approach, allowing bang gap tuning over a broad spectral range via a combination of size effects and halide stoichiometry.

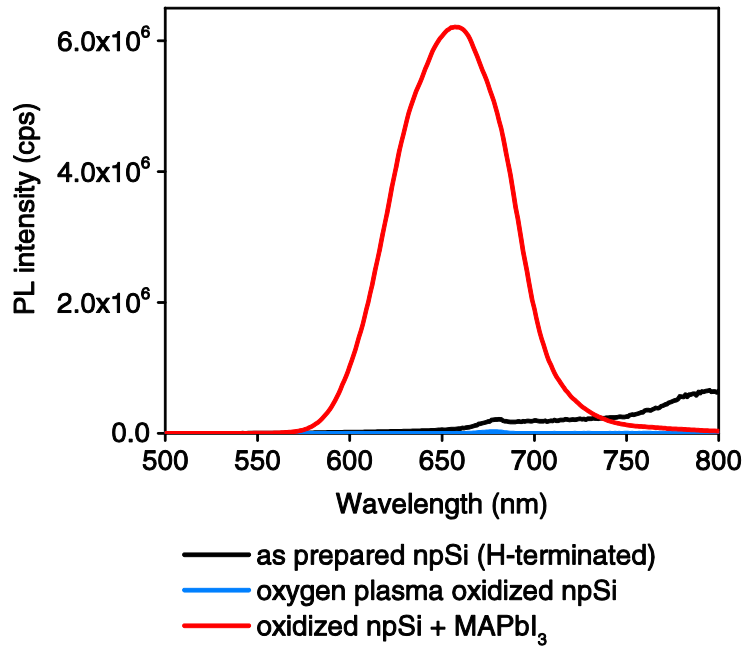

**fig. S2. PL of native npSi and MAPbI3 nanocrystals in npSi.** The as-prepared npSi surface is hydrogen-terminated and shows weak photoluminescence in the NIR (black line). A short oxygen plasma treatment (5 min at 50 W) oxidizes the npSi surface, resulting in complete photoluminescence quenching (blue line). Subsequent infiltration with MAPbI$_3$ precursor solution yields nanocrystalline MAPbI$_3$ exhibiting PL emission (red line) that is strongly blue shifted compared to bulk films.

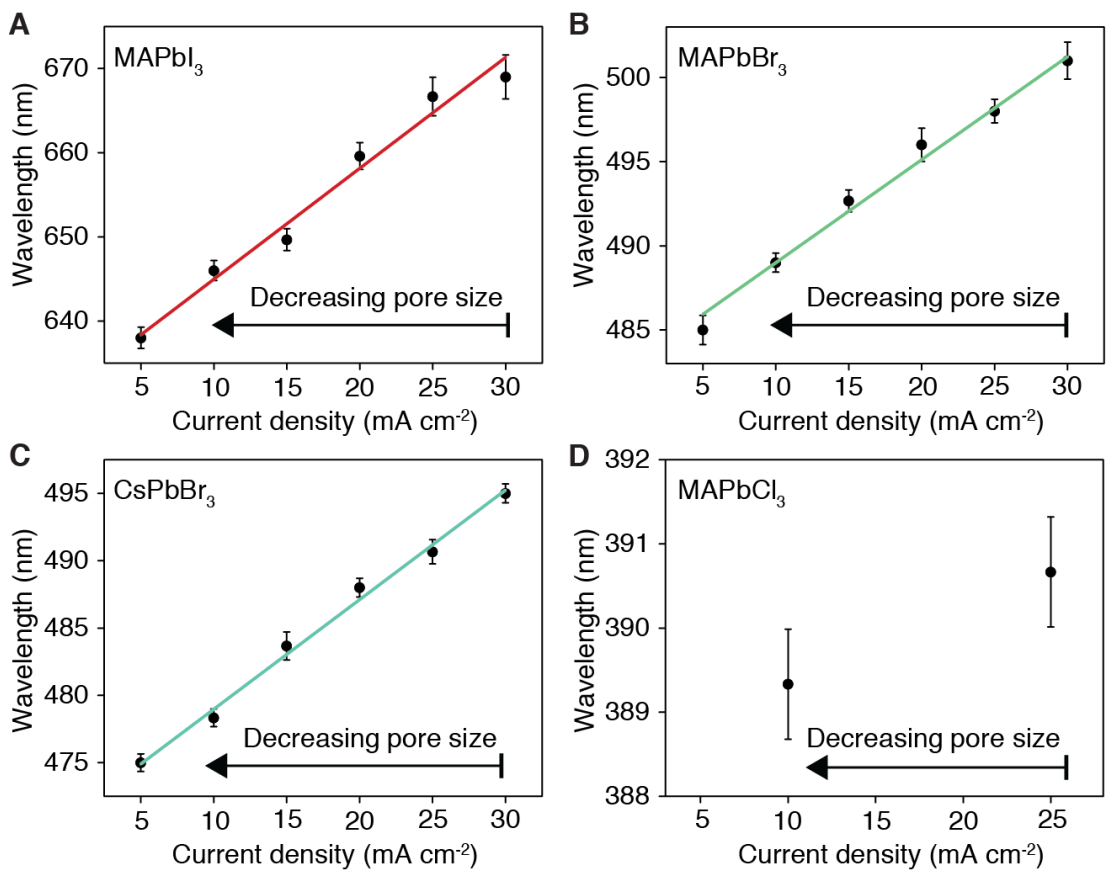

**fig. S3. Fine-tuning of the bandgap in npSi via variation of pore size.** The pore size in anodic npSi depends on the anodization current density provided by the doped parent wafer. A decrease in current density corresponds to decreasing pore sizes *(42, 63, 64)*. The PL peak emission in ncAPbX$_3$ is precisely fine-tuned by variation of npSi matrix anodization parameters. We investigated current densities from 5 mA cm$^{-2}$ up to 30 mA cm$^{-2}$ using low-doped p-type Si wafers. A linear dependence of PL peak position on the anodization current is observed for **(A)** MAPbI$_3$, **(B)** MAPbBr$_3$, **(C)** CsPbBr$_3$, and **(D)** MAPbCl$_3$. Black data points and error bars represent mean value and standard deviation over at least four individual samples each. Linear fits of the data are shown as colored lines.

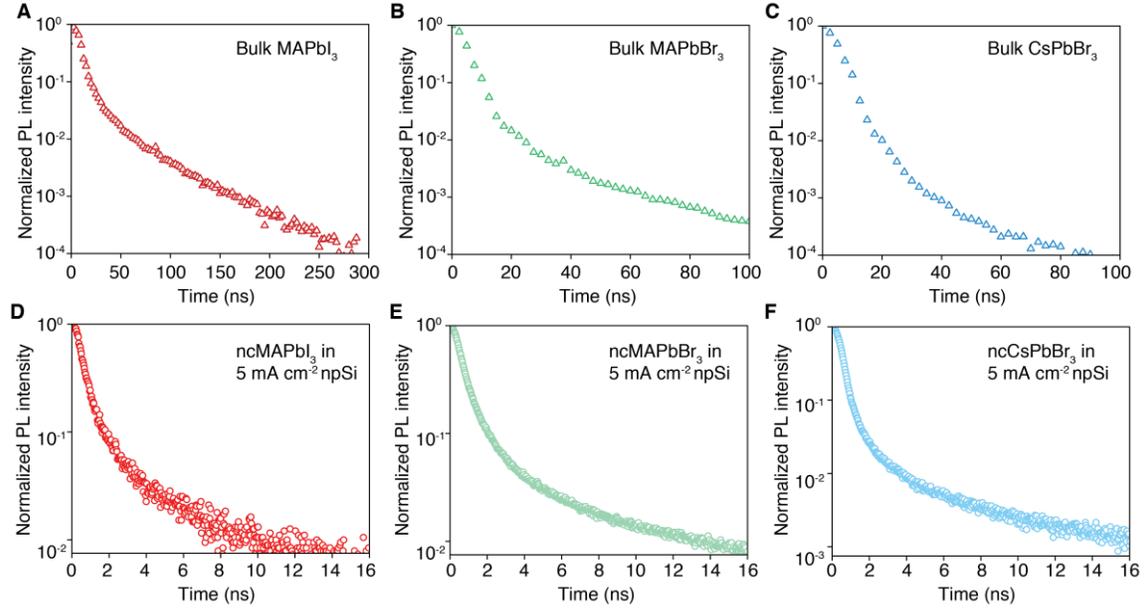

**fig. S4. Time-resolved PL lifetime.** Bulk films of **(A)** $MAPbI_3$, **(B)** $MAPbBr_3$ and **(C)** $CsPbBr_3$ spin-cast on glass slides from undiluted parent solutions typically show a bi-exponential decay. We here record the full emission spectrum (gate step 2.5 ns, gate width 2.5 ns) following excitation, and integrate over the baseline corrected peak to extract the time resolved PL (open triangles in A-C). We find time constants of $\tau_1 = 4.8$ ns and $\tau_2 = 25.4$ ns for $MAPbI_3$, $\tau_1 = 3.4$ ns and $\tau_2 = 12.0$ ns for $MAPbBr_3$ and $\tau_1 = 3.7$ ns and $\tau_2 = 8.3$ ns for $CsPbBr_3$. Corresponding nanocrystals grown in 5–mA cm$^{-2}$ npSi films show shorter lifetimes, a trend that is typically observed for quantum dots *(15, 44)* and other nanosized perovskite emitters *(27, 45)*. Here we find a bi-exponential decay behavior for **(D)** $ncMAPbI_3$ with $\tau_1 = 0.4$ ns and $\tau_2 = 2.2$ ns, **(E)** $ncMAPbBr_3$ with $\tau_1 = 0.5$ ns and $\tau_2 = 1.8$ ns and **(F)** $ncCsPbBr_3$ with $\tau_1 = 0.3$ ns and $\tau_2 = 3.1$ ns.

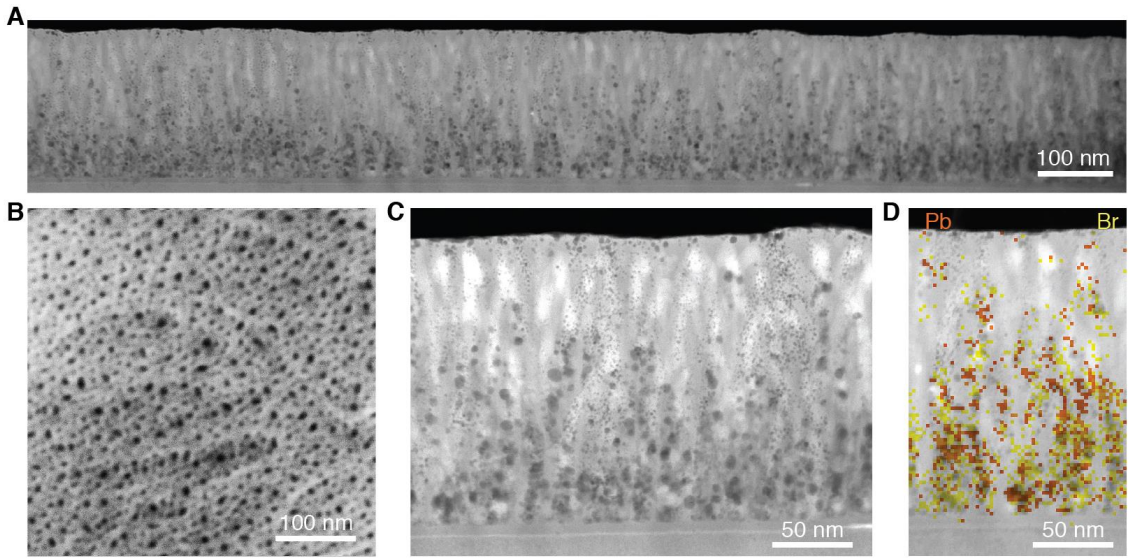

**fig. S5. Structural characterization of MAPbBr3 nanocrystals in npAAO.** (**A**) STEM image of a 170 nm high, ~100 nm thick npAAO filament prepared by FIB milling, partially-filled with perovskites. Infiltration of the ~170 nm long oxide nanotubes with diameter of ~6-8 nm is achieved across the whole surface area. The highest density of perovskite is observed predominantly at the bottom of the pores. The nucleation density and thus the apparent filling height is controllable via precursor concentration, with this we achieve filling of several 100 nm high npAAO scaffolds. (**B**) Surface SEM of npAAO fabricated on glass substrates. A porosity of ~15 % is calculated from these images. (**C**) Detailed view of the BF-STEM projection with leadrich perovskite appearing dark. (**D**) EDX map of Pb (orange) and Br (yellow) species overlaid onto the BF-STEM image, indicating successful formation of MAPbBr$_3$ nanocrystal domains in the npAAO template.

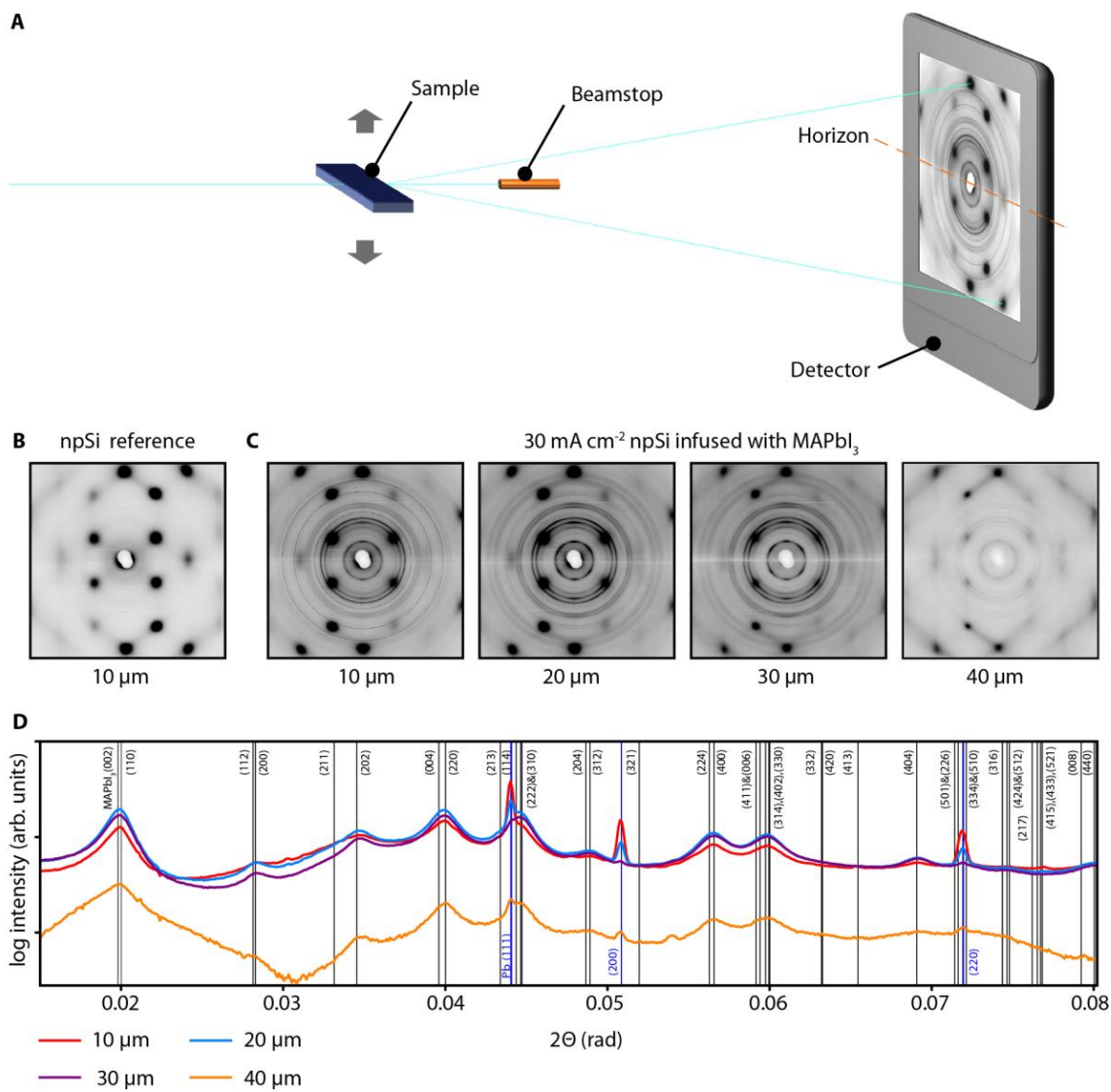

**fig. S6. Wide-angle x-ray scattering (WAXS) study of MAPbI₃ nanocrystals in 30–mA cm⁻² npSi: Experiment and data analysis. (A)** Measurement geometry of the high energy x-ray depth profiling experiment. The sample surface was thoroughly aligned parallel to the microfocus x-ray beam to achieve beam size limited resolution (vertical FWHM: 5 µm) for depth scanning. A beamstop was placed behind the sample to protect the detector from the transmitted beam and to reduce air scattering. The sample detector distance was 2000 mm. All measurements shown in this figure (sample batch 1) were performed with an x-ray energy of 98.5 keV. **(B)** Reference diffraction pattern of a npSi layer without perovskite. The regularly- arranged broad peaks

correspond to the crystal structure of oriented Si *(65)* penetrated along the [110] direction. Such reference patterns were recorded at different depths below the surface and were used to identify and remove the background signal from the measurements on MAPbI$_3$ containing npSi layers. **(C)** Diffraction patterns of a 30–mA cm$^{-2}$ npSi layer infused with MAPbI$_3$ probed at different depths below the surface. The images are a superposition of the background from the porous Si (cf. fig. S6B) and continuous broad Debye rings from the perovskite crystallites within the pores. Close to the surface additional sharp inhomogeneous Debye rings appear, which can be attributed to traces of metallic Pb. The raw data were processed using the "Nika" package *(60)* for Igor Pro (WaveMetrics). The peaks resulting from Si were masked and the azimuthal average of the remaining area was calculated. Subsequently the signature of the porous Si (cf. fig. S6B) at the corresponding depth was subtracted. **(D)** Background corrected, azimuthally averaged intensity profiles of MAPbI3 nanocrystals in 30–mA cm$^{-2}$ npSi at different depths below the surface. The peaks are indexed to a tetragonal structure with the lattice constants a = b = 8.90(3) Å and c = 12.71(5) Å (black vertical lines). Pb peaks are marked by blue vertical lines. To estimate the depth-dependent crystallite size Λ of the perovskite from the FWHM of the peaks, we use the Scherrer equation *(48)*: $FWHM = 2 \times \sqrt{\frac{\ln 2}{\pi}} \times \frac{\lambda}{\Lambda} \times \frac{1}{\cos(\frac{2\Theta}{2})}$. Here λ = 0.1259(3) Å is the wavelength and 2Θ is the scattering angle. The ranges around three peaks consisting of two MAPbI3 reflections each ((002)&(110), (004)&(220) and (224)&(400)) were fitted by a superposition of six Gaussian peaks with fixed peak positions and FWHM as expected from the Scherrer equation with Λ as fit parameter.

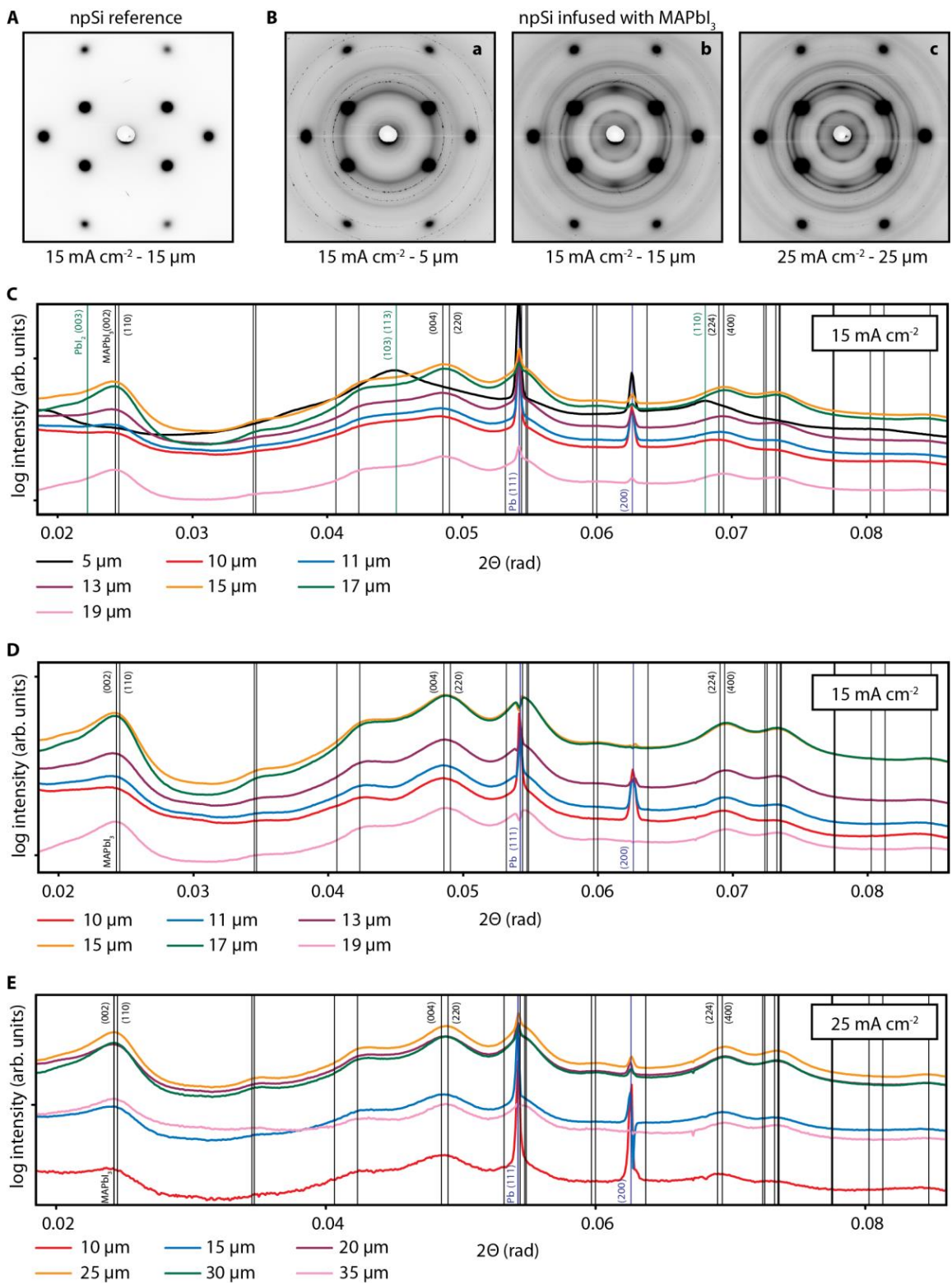

**fig. S7. WAXS study of MAPbI₃ nanocrystals in 15–mA cm⁻² npSi and 25–mA cm⁻² npSi.**

These samples are part of batch 2 and were measured in a second experimental session. The measurement geometry was identical to that described in fig. S6A, but the x-ray energy was 80.0 keV and the vertical FWHM of the beam was 2 µm. **(A)** Reference diffraction pattern of a 15–mA cm$^{-2}$ npSi layer without perovskite, revealing the signal of the porous Si. **(B)** Diffraction patterns of a 15–mA cm$^{-2}$ npSi layer and a 25–mA cm$^{-2}$ npSi layer infused with MAPbI$_3$. All diffraction patterns show the contributions seen in fig. S6C. Primarily close to the surface (5 µm) additional Debye rings appear, which can be ascribed to PbI$_2$ *(66)*. **a** 15–mA cm$^{-2}$ npSi/ncMAPbI$_3$ measured 5 µm below the surface. This measurement shows the most pronounced PbI$_2$ signature. **b** 15–mA cm$^{-2}$ npSi/ncMAPbI$_3$ probed at the depth of maximum perovskite signal (15 µm below the surface). **c** 25–mA cm$^{-2}$ npSi/ncMAPbI$_3$ probed at the depth of maximum perovskite signal (25 µm below the surface). (C) Azimuthally averaged intensity profiles of MAPbI$_3$ in 15–mA cm$^{-2}$ npSi at different depths. The background of the npSi has been removed as described in fig. S6C. Vertical lines mark the calculated peak positions for the diffraction signal of the tetragonal MAPbI3 crystallites (black) (47) and crystalline residues of PbI$_2$ (green, only the most intense peaks are marked) *(66)* and Pb (blue). The diffraction pattern recorded 5 µm below the surface (black curve) is dominated by the signature of the residues and can be used to remove their influence from the other patterns: For each depth, the contribution of PbI$_2$ to the diffraction profile was quantified by comparing the intensity at the position of the most intense PbI$_2$ diffraction peak ((103)&(113)) with the corresponding intensity value at the unmodified high angle flank of the MAPbI$_3$ (004)&(220) diffraction peak. The respective proportion of the PbI$_2$ signal - represented by the diffraction profile recorded 5 µm below the surface - was then subtracted from the intensity profile to isolate the contribution of the perovskite crystallites. **(D to E)** The resulting isolated MAPbI$_3$ diffraction profiles for 15–mA cm$^{-2}$ npSi/ncMAPbI3 **(D)** and 25–mA cm$^{-2}$ npSi/ncMAPbI3 **(E)**. From these curves the crystallite sizes were estimated following the procedure described in fig. S6D with the difference that the wavelength was λ=0.1550(3) Å and the peak positions were directly calculated from the structure described by

Stoumpos *et al.(47)* (tetragonal with a = b = 8.849(2) Å and c = 12.642(2) Å), as no deviation from this structure was observed.

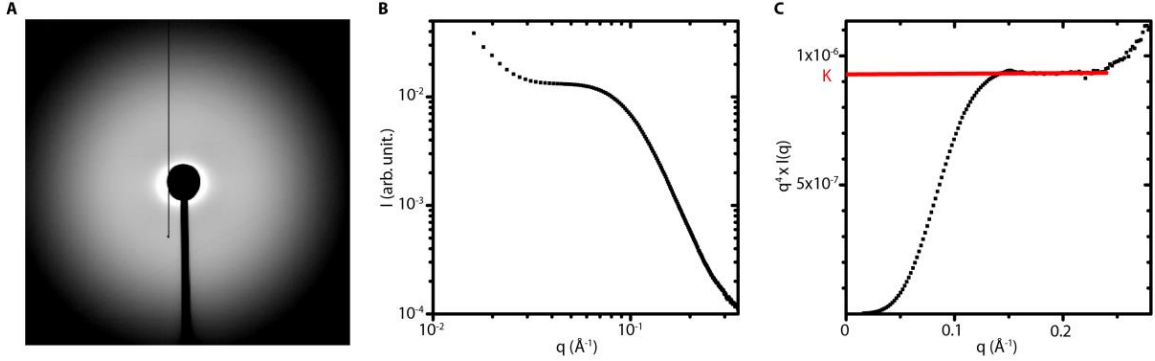

**fig. S8. Pore size estimation by Porod analysis of small-angle x-ray scattering (SAXS) data of 15–mA cm$^{-2}$ npSi.** (A) SAXS intensity pattern from 15–mA cm$^{-2}$ npSi recorded in transmission with the sample surface perpendicular to the 20.0 keV x-ray beam. For data reduction the intensity was averaged in azimuthal direction using the "Nika" package *(60)* for Igor Pro (WaveMetrics). (**B**) Azimuthally averaged intensity profile of 15–mA cm$^{-2}$ npSi plotted against the scattering vector **q** in double logarithmic representation, where $q = \frac{4\pi}{\lambda} \sin\left(\frac{2\Theta}{2}\right)$, $2\Theta$ is the scattering angle, and $\lambda$ is the incident x-ray wavelength. The scattering signal of the nanopores is dominating in the region of 0.05 Å$^{-1}$ ≤ q ≤ 0.24 Å$^{-1}$. (**C**) Porod representation $I(q) \times q^4$ vs. $q$ of the same data. The curve displays a plateau, which can be extrapolated to zero-q to give the so called Porod constant K as the ordinate intercept. K is proportional to the summed surface area of all pores in the beam *(51)*. The Porod invariant Q is defined by $Q = \int_0^\infty q^2 I(q) dq$ nd was obtained by numerically integrating the extrapolated scattering curve. The porosity Φ was determined from the x-ray transmission T$_{npSi}$ of the porous layer and the transmission T$_{Si}$ of the bulk Si, which were measured with a microfocus (2 µm FWHM) high energy x-ray beam (80.0 keV) parallel to the sample surface: $\Phi = 1 - \frac{\ln T_{npSi}}{\ln T_{Si}} = 57\%$. The average pore diameter d was estimated by $d = \frac{4}{\pi(1-\Phi)} \times \frac{Q}{K} = 40$ Å *(67)*.